\documentclass[twocolumn]{article}

\usepackage[left=1in,right=1in,top=1in,bottom=1in]{geometry}
\usepackage{tikz}
\usetikzlibrary{calc,patterns,angles,quotes}

\usepackage{algorithm}
\usepackage{algpseudocode}
\usepackage{amsmath}
\usepackage{authblk}
\usepackage{url}

\usepackage[explicit]{titlesec}
\renewcommand{\thesection}{\Roman{section}}
\renewcommand{\thesubsection}{\thesection.\Alph{subsection}}
\renewcommand{\thesubsubsection}{\thesubsection.\arabic{subsubsection}}

\titleformat{\section}{\large\scshape\centering}{\thesection.\space}{0pt}{#1}[]
\titlespacing*{\section}{0pt}{0.5\baselineskip}{0pt}
\titleformat{\subsection}{\normalsize\itshape}{\Alph{subsection}.\space}{0pt}{#1}[]
\titlespacing*{\subsection}{0pt}{0.5\baselineskip}{0pt}
\titleformat{\subsubsection}{\normalsize\itshape}{\arabic{subsubsection}.\space}{0pt}{#1}[]
\titlespacing*{\subsubsection}{0pt}{0.5\baselineskip}{0pt}
\usepackage{geometry}
\newgeometry{left=2cm, right=2cm, top=2.5cm,bottom=3.5cm}
\makeatletter
\renewcommand{\fnum@figure}{Fig. \thefigure}
\renewcommand{\fnum@table}{Tab. \thetable}
\makeatother

\usepackage{nopageno}
\renewcommand{\abstractname}{}

\title{\vspace{-1.0cm}\textbf{\normalsize NeuralODEs for VLEO simulations: Introducing thermoNET for Thermosphere Modeling}}

\author[1]{Dario Izzo} 
\author[1,2]{Giacomo Acciarini} 
\author[3]{Francesco Biscani}

\affil[1]{\emph{Advanced Concepts Team, European Space Research and Technology Centre (ESTEC), Noordwijk, The Netherlands.}}
\affil[2]{\emph{Surrey Space Centre, University of Surrey, Guildford, United Kingdom}}
\affil[3]{\emph{ESA/ESOC, Darmstadt, Germany}}

\date{}  
\begin{document}

\maketitle

\begin{abstract}
\vspace{-1.15\baselineskip}
\textbf{\emph{\quad Abstract} - We introduce a novel neural architecture termed thermoNET, designed to represent thermospheric density in satellite orbital propagation using a reduced amount of differentiable computations. 
Due to the appearance of a neural network on the right-hand side of the equations of motion, the resulting satellite dynamics is governed by a NeuralODE, a neural Ordinary Differential Equation, characterized by its fully differentiable nature, allowing the derivation of variational equations (hence of the state transition matrix) and facilitating its use in connection to advanced numerical techniques such as Taylor-based numerical propagation and differential algebraic techniques. 
Efficient training of the network parameters occurs through two distinct approaches.
In the first approach, the network undergoes training independently of spacecraft dynamics, engaging in a pure regression task against ground truth models, including JB-08 and NRLMSISE-00. In the second paradigm, network parameters are learned based on observed dynamics, adapting through ODE sensitivities. In both cases, the outcome is a flexible, compact model of the thermosphere density greatly enhancing numerical propagation efficiency while maintaining accuracy in the orbital predictions.}
\end{abstract}

\section{Introduction}
The analysis of satellite trajectories, particularly when influenced by the upper atmosphere, is affected by critical uncertainties. While precise models of our thermosphere such as the widely used NRLMSIS from the United States Naval Research Laboratory~\cite{picone2002nrlmsise}, the Jacchia-Bowman (JB)~\cite{bowman2008new}, the Drag Temperature Models (DTM)~\cite{bruinsma2015dtm}, and the High Accuracy Satellite Drag Model (HASDM)~\cite{storz2005high}, have become standard in orbital simulations, their significant discrepancies underline the challenge of accurately estimating drag effects. 
In a recent work by Yin et al.~\cite{yin2022evaluation} quantitative measures of the relative errors introduced by empirical atmospheric models are presented using as ground truth the density inferred in the thermosphere from Swarm-C \cite{friis2008swarm} orbital data and show relative deviations as large as 300\% and overall mostly larger than 20\%. 
Inherent uncertainties in established thermosphere models are thus significant, suggesting that artificial neural modelling may be particularly appropriate in this context. It is thus not surprising that the idea of using an artificial neural network to represent the density in the thermosphere has been introduced and studied in the recent past in several different works~\cite{perez2015neural, licata2021improved, cui2021atmospheric, acciarini2024improving}, producing a plethora of neural models able to directly predict the density in the thermosphere or to correct existing models.\newline
In this work, we introduce a novel neural architecture, an implicit neural representation \cite{mildenhall2021nerf} we call thermoNET which is revealed to be able to achieve high accuracy in reproducing thermospheric empirical models while only using a limited number of learnable parameters.
The compact size of the resulting networks is essential for enabling their utilization in space flight mechanics applications, including orbital decay studies, re-entry analysis, and long-term debris population analysis. The guaranteed asymptotic behaviour of thermoNETs (as the altitude grows to infinity) is also an important property that avoids numerical issues when using the model outside its training bounds.
Moreover, the inherent differentiability of our neural representation allows the automated derivation of variational equations, hence of the state transition matrix, and facilitates its integration with modern numerical techniques, such as Taylor numerical integration \cite{biscani2021revisiting}, differential algebra tools \cite{armellin2010asteroid}, and the neuralODEs approach \cite{chen2018neural}, thus offering additional advantages in terms of computational efficiency and overall capabilities of orbital propagation.

\section{Results}
\subsection{thermoNETs: a differentiable atmospheric model}
A thermoNET takes geodetic coordinates of a point in the thermosphere and selected space weather indices as inputs, yielding the corresponding air density at that location. 
\begin{figure}[tb]
  \centering
    \includegraphics[width=0.5\textwidth]{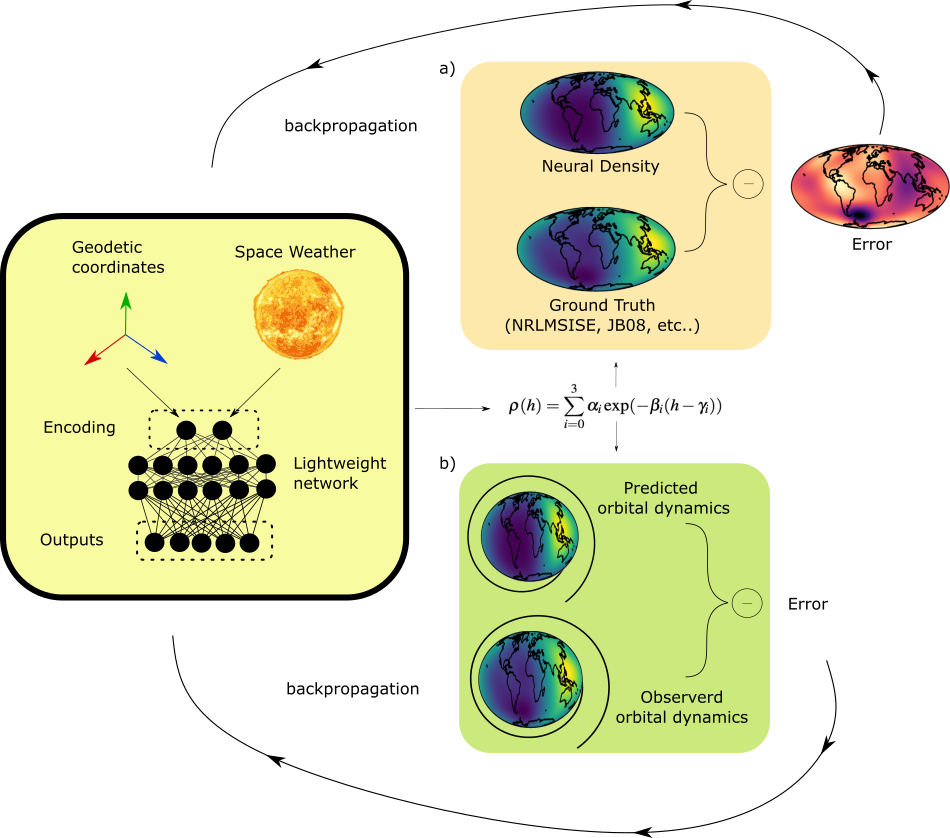}
  \caption{Thermonets at a glance. A lightweight artificial neural network transforms geodetic coordinates and space weather data into the predefined air density model. The network parameters are learned via backpropagating the error via two distinct learning pipelines: a) to match some ground truth model (regression learning) or b) to match orbital observations (neural ODE learning).
  \label{fig:thermonets}}
\end{figure}
Our proposed architecture, illustrated in Fig.~\ref{fig:thermonets}, employs a straightforward linear combination of exponential terms reminiscent of isothermal altitude trends. 
The adjustment of the model parameters to account for latitude, longitude, and solar weather effects is delegated to a lightweight feed-forward neural network.
Recent advancements in neural modeling, such as Neural Implicit Representations \cite{mildenhall2021nerf}, Physical Informed Neural Networks \cite{cai2021physics}, and DeepONets \cite{lu2021learning}, have demonstrated the impressive representation power of neural models in real-world applications. 
These approaches extend beyond universal approximation theorems, providing insights into the necessary network size and complexity. 
Building on these ideas, our thermoNETs incorporate certain aspects of thermosphere physics directly, but they achieve this without relying on integro-differential operators.
Instead, our approach involves forcing a functional form, related to underlying physical models, where the important parameters are then represented neurally.
The incorporation of physics into the resultant model distinguishes thermoNETs from some of the prior works exploring neural models in analogous contexts \cite{perez2015neural, licata2021improved, cui2021atmospheric, acciarini2024improving}. 
Most of these prior efforts have yielded more intricate architectures, applicable within narrow altitude ranges, and lacking guarantees on the asymptotic behavior of the represented density. These limitations restrict substantially their broader applicability.
In contrast, our thermoNETs favor simplicity, utilizing fewer than a few thousand parameters and guaranteeing vanishing values for the density $\rho$ as the altitude approaches infinity.
They demonstrate efficacy across a broad altitude spectrum, spanning from 180 to well beyond 1000 km, and exhibit relative errors limited to a few percentage points when compared to the ground truths they seek to reproduce.

\begin{figure*}[tb]
  \centering
    \includegraphics[width=\textwidth]{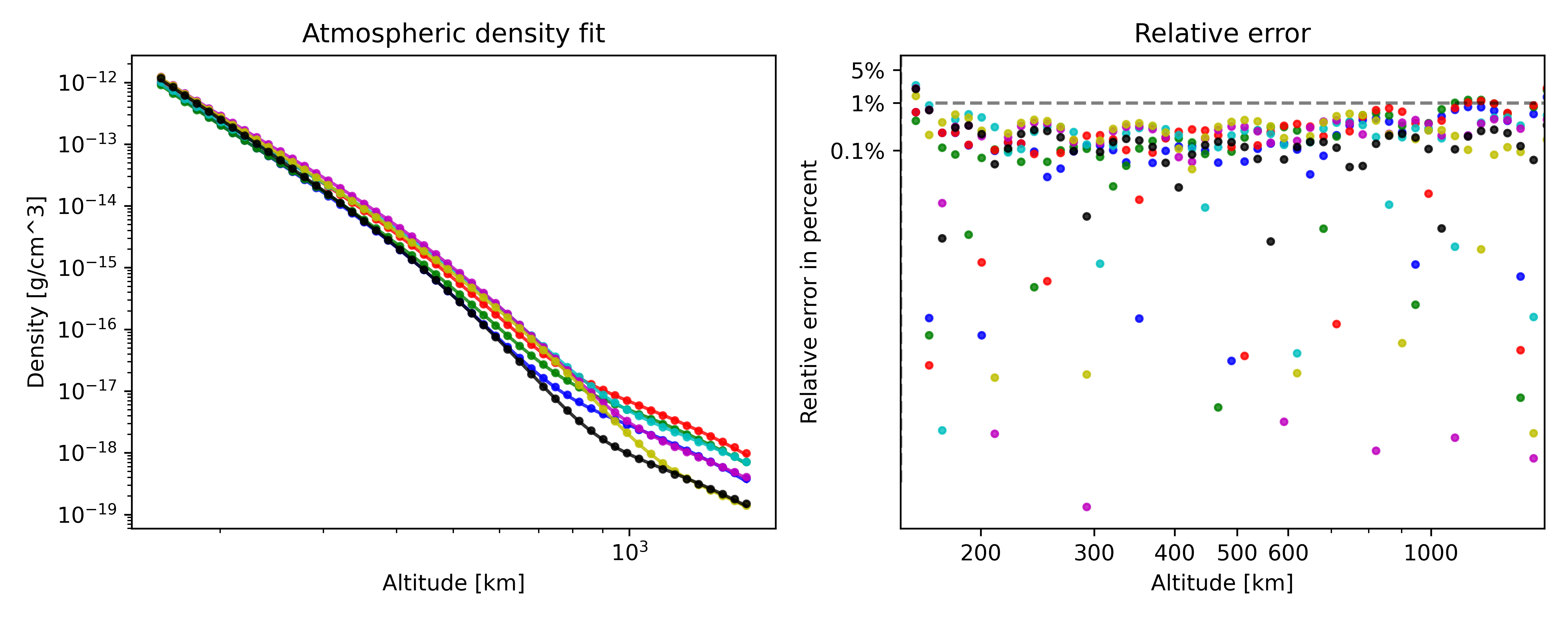}
  \caption{Left: Ground truth (NRLMSISE-00), and model predictions fitted assuming a fixed longitude, latitude, and solar weather. Each case corresponds to an independent fit of the coefficients in Eq.~\eqref{eq:rho_functional_form} and the continuous line represents the ground truth, while the overlaid markers represent the fits. Right: relative error introduced by the fits. All interesting cases (above 180km) are mostly well below a 1\% error as highlighted by the dotted line.
  \label{fig:rho_error}}
\end{figure*}

Some aspects of the thermosphere physics are included explicitly by assuming the following functional form for the density as a function of the altitude $h$:
\begin{equation}
\label{eq:rho_functional_form}
\rho(h) = \sum_{i=0}^3\alpha_i\exp(-\beta_i(h-\gamma_i))
\end{equation}
where the sum of exponential terms resembles isothermal contributions. The chosen functional form allows it to fit well, at fixed longitudes, latitudes, and space weather parameters, the variation of the air density with the altitude in the thermosphere as highlighted in Fig.~\ref{fig:rho_error}. 
In the figure we have generated data using the NRLMSISE-00 model and obtained, at several values of geodetic longitude, latitude, and space weather parameters, the values $\rho_N$ of the Earth's atmospheric density from 160 to 1600 km. 
In each case, we fitted the functional form in Eq.~\eqref{eq:rho_functional_form} and plotted it against the ground truth together with the relative error introduced by the fit defined as $\mbox{err} = \frac{|\rho - \rho_{N}|}{\rho_{N}}$.
Starting from altitudes larger than 180km the discrepancy with the NRLMSISE-00 model is well below 1\% most of the time. 

A thermoNET models the atmospheric density using Eq.~\eqref{eq:rho_functional_form}, where the coefficients $\alpha_i, \beta_i$ and $\gamma_i$ are constructed from the output of a small feed forward neural network $\mathcal N_\theta(\lambda, \varphi, DOY, SID, \mathcal S_w)$ receiving the geodetic coordinates $\lambda, \varphi$ as input (see Appendix \ref{app:geodetic} for a definition of a differentiable geodetic coordinate transformation), as well as the day of year (DOY), the seconds in day (SID) and space weather parameters, here indicated generically with $\mathcal S_w$. 

\begin{figure*}[p!]
  \centering
    \includegraphics[width=0.9\textwidth]{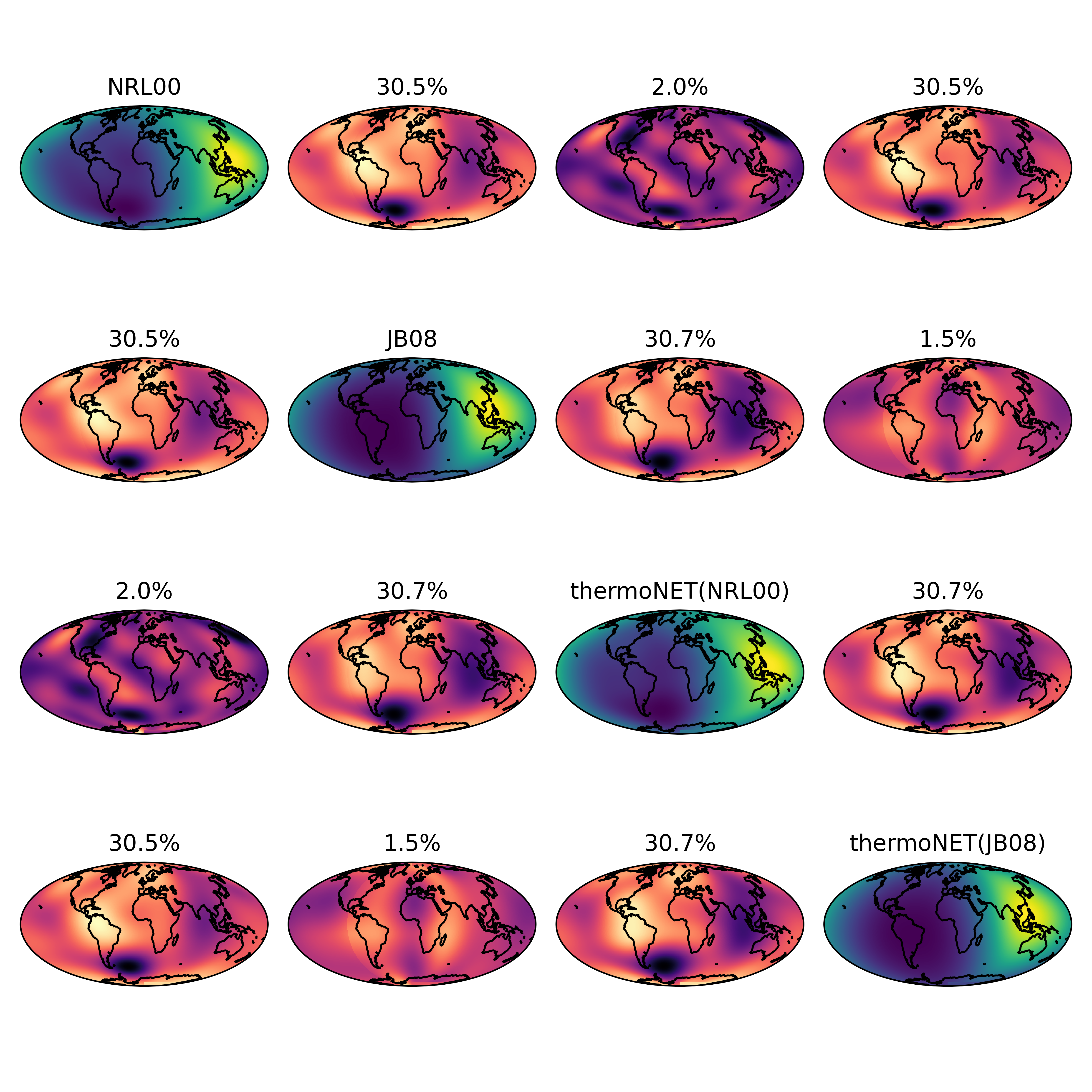}
  \caption{Prediction (diagonal) and distribution of the absolute percentage difference between models (off-diagonal) at an altitude of 400 km and on the 22nd of April 2018 at 5:13:35 (GMT). The mean is also indicated. The ability of the neural models to approximate well atmospheric trend is evident as the error associated is one order of magnitude smaller than the difference between models.
  \label{fig:pairplots}}
\end{figure*}

\subsection{Datasets}
While in this study we primarily focus on the widely used empirical models NRLMSISE-00 (NRL00)~\cite{picone2002nrlmsise} from the United States Naval Research Laboratory and the Jacchia-Bowman \cite{bowman2008new} (JB08) model as target densities, our approach can equivalently be used with any other data source. Both models, to compute a density $\rho$ in a point of the thermosphere, require the geodetic coordinates $h, \lambda, \varphi$, the epoch and space-weather information denoted by the symbol $\mathcal S_w$. We generate two comprehensive databases of air densities $\mathcal D_{NRL00}$ and $\mathcal D_{JB08}$ using both these empirical models.
To construct the databases, we sample the geodetic longitude $\lambda$ and geodetic latitude $\varphi$ on a uniform 100x100 grid. For each obtained point, we randomly select an epoch within the years 2009-2022 and extract the corresponding space weather data as provided by the US National Oceanic and Atmospheric Administration (NOAA). Subsequently, utilizing the values of $\lambda, \varphi$, and $\mathcal S_w$, we compute the density $\rho$ at 100 altitudes within the range of $180$ to $1000$ km, sampled at 100 points on a logarithmic scale as to favor low altitudes. The resulting databases contain 1 million points, representing a diverse range of epochs, and space weather conditions, and covering the selected altitude range globally.
To provide an indication of the accuracy expected when utilizing empirical models to predict density values within the range considered, we report the average relative difference between JB08 and NRLMSISE-00 on the constructed dataset to be approximately 72\%.

\subsection{Learned models}
As previously noted, the dependence of the density on altitude is well captured by Eq.~\eqref{eq:rho_functional_form}, which is then enforced in the thermoNET modelling. 
We thus train two feed-forward neural networks $\mathcal N_{NRL00}$ and $\mathcal N_{JB08}$ on each of the two datasets produced to predict the values of $\alpha_i, \beta_i$ and $\gamma_i$, which are then fed into Eq.~\eqref{eq:rho_functional_form} to compute a predicted density $\hat \rho$ and hence the loss $\mathcal L = \frac{100}{N}\sum_{i=1}^N |\rho_i - \hat\rho_i|/ \rho_i$. Details on the exact definitions of the attributes used and the network outputs, as well as details on the training procedure, can be found in the Appendix \ref{sec:training}.
Both networks, as outlined in Tab.~\ref{tab:results}, share a similar structure, differing only in the dimension of the input layer which is aligned with the specific space weather indices considered by the corresponding empirical model. With fewer than a few thousand parameters (roughly two orders of magnitude less than comparable models in \cite{licata2021improved}), they adeptly approximate the target density with errors limited to a few percentage points. 

\begin{table*}[tb]
\begin{center}
\begin{tabular}{ c|c|c|c|c|c } 
  & n. parameters & average rel. err ($\mathcal L$) & max rel. err & layers dim& dataset\\ 
  \hline
 $\mathcal N_{NRL00}$ & 1804 & \textbf{2.17}\% & 32.93\% & 10x32x32x12&$\mathcal D_{NRL00}$\\ 
 $\mathcal N_{JB08}$ & 1996 & \textbf{1.43}\% &21\% &16x32x32x12&$\mathcal D_{JB08}$\\ 
 $\mathcal K_{S}$ & 1977 & 3.88\%  & 78.18\%&11x38x38x1&$\mathcal D_{NRL00}$\\ 
 $\mathcal K_{B}$ & 257,001 &  0.98\% & 15.45\% &11x500x500x1&$\mathcal D_{NRL00}$\\ 
\end{tabular}
\end{center}
\caption{Details on the thermoNETs developed ($\mathcal N_{NRL00}$, $\mathcal N_{JB08}$) and 
networks directly predicting the density as output ($\mathcal K_{S}$, $\mathcal K_{B}$), thus not enforcing Eq.~\eqref{eq:rho_functional_form}\label{tab:results}.}
\end{table*}

As a first comparison, we utilized our datasets to train two additional neural architectures, denoted as $\mathcal K_s$ and $\mathcal K_b$, directly predicting the density through a feed-forward neural network (FFNN), akin to the approach employed in the thermosphere neural models implemented in the open-source Karman software\footnote{\url{https://github.com/spaceml-org/karman}}, developed by Trillium Technologies in collaboration with the Heliophysics Division of NASA. 
We trained a small network, matching the number of parameters of thermoNETs experiments, as well as the training setup: we used in both cases 2,000 epochs, decreasing the learning rate from $10^{-3}$ to $10^{-4}$ after 1,000 epochs. Moreover, we also trained a much larger network, mirroring the dimensions of the bigger ones presented in~\cite{acciarini2024improving}. 
In the latter case, as expected, we were able to surpass for example, the performance levels of $\mathcal N_{NRL00}$ as it achieved final relative percentage errors of about 0.98\%. However, and most importantly, in the other case a significant deterioration of the mean average percentage error was observed (from roughly 2.17\% of the physics-informed structure to 3.88\%), confirming the intuition that informing the network with the physics model in Eq.~\eqref{eq:rho_functional_form} helps with the model accuracy, allowing the training and network parameters to ignore the altitude and focus solely on reproducing the other dependencies.

A further test was made comparing the networks inference time to the computational time of the original empirical atmospheric models. 
Without going into much details, we report that in all tests performed, we observed how the neural models consistently offered a significant speed increase (largely exceeding a factor two). 
The improvement in computational speed is to be attributed to the network's small size resulting in a significant compression of the original models. 

Finally, we validate our models on test data generated independently from the datasets $\mathcal D$. To do so we tested them at specific altitudes, epochs, and space weather conditions computing the quality of the reconstructed thermospheric density over the whole range of latitudes and longitudes. In Fig.~\ref{fig:pairplots}, we show the results for the case of an altitude of 400 km and the epoch 22nd of April 2018, 5:13:35 (GMT). The high quality of the reconstructed densities is confirmed, as well as large differences between empirical models. 

\section{Exploiting differentiability}
A thermoNET possesses distinctive properties that set it apart from prior attempts at neural modeling of the thermosphere.
These include its compact size, its physically informed nature (as evidenced by Eq.~\eqref{eq:rho_functional_form}), and the guarantees on its asymptotic behavior. 
These properties, coupled with the inherent differentiability of the model, open up avenues for applications that would otherwise be unattainable or less effective. 
In the subsequent two sections, we provide brief introductions to two such applications and present preliminary demonstrations of the anticipated advantages.

\begin{figure*}[tb]
  \centering
    \includegraphics[width=\textwidth]{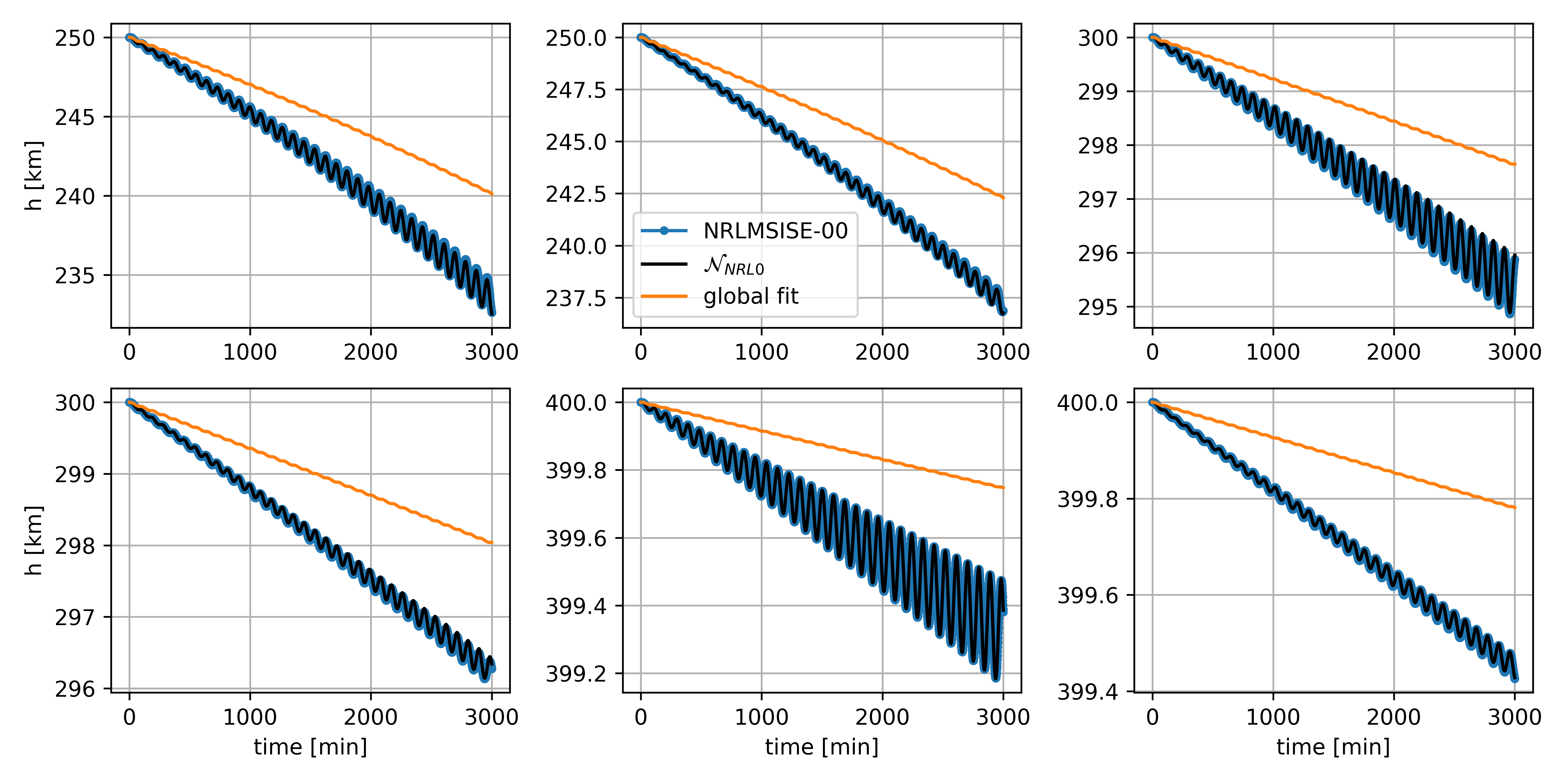}
  \caption{Altitude trends along various numerical propagations of VLEO satellites where the thermospheric density is represented by a) the NRLMSISE-00 empirical model, b) the thermoNET $\mathcal N_{NRL00}$, or c) a fit accounting only for altitude dependence. For all practical purposes the thermoNET yields equivalent numerical results.
  \label{fig:orbits}}
\end{figure*}

\subsection{Taylor integration}
The Taylor integration technique \cite{jorba2005software, scott2010high} stands out as one of the most efficient numerical schemes for orbital propagation. Recent advancements in computer implementations \cite{biscani2021revisiting, perez_hernandez_2019_2562353} have largely overcome the challenges associated with its use, automating the manipulations required to develop such integrators. 
However, the technique still encounters limitations in cases prevalent in space flight mechanics where non-differentiable terms appear on the right-hand side of the differential equations\footnote{It is worth noting that from a strictly mathematical point of view the notion of differentiability, in the systems here considered, is generally satisfied almost everywhere. We here consider differentiability also from a computational standpoint. This pertains to the feasibility of natively performing automated differentiation on space flight mechanics code.}.
In instances where non-differentiable terms are present, a solution to allow the use of the Taylor method involves creating a neural representation of the non-differentiable terms and leveraging it to construct the Taylor integration scheme. 
This approach was introduced in \cite{biscani2022reliable}, where a Taylor integrator with event detection was developed for dynamics incorporating complex geometrical eclipse conditions.
Similarly, the thermoNETs developed in this study facilitate the incorporation of the non-differentiable NRLMSISE-00 and JB08 models.
To illustrate the advantages, let us consider a satellite undergoing Keplerian motion perturbed by thermospheric drag. Its dynamics is described by:

\begin{equation}
\label{eq:neuralODE}
\dot{\pmb v} = -\frac{\mu}{r^3} \pmb r - \frac 12  \frac{\rho}{C_b} v_{r}\pmb v_r
\end{equation}
where the thermospheric density is indicated by $\rho$, the ballistic coefficient with $C_b$ and the spacecraft relative velocity to the thermosphere with $v_r$. For our benchmark, we will ignore complicated wind profiles and thus set:
$$
\pmb{v}_r=\dfrac{d\pmb{r}}{dt}-\pmb{\omega}\times \pmb{r}=\bigg[\dfrac{dx}{dt}+\omega_z y, \ \dfrac{dy}{dt}-\omega_z x, \ \dfrac{dz}{dt}\bigg]^T
$$
where $\pmb{\omega}$ is the Earth's angular velocity around its spin axis. Using these equations of motion, we perform several numerical orbital propagations of satellites in VLEO  computing the thermospheric density $\rho$ using the NRLMSISE-00 empirical model as well as $\mathcal N_{NRL00}$. The VLEO are assumed circular and have starting altitudes of $250, 300$ and $400$ km. For each altitude, we consider both a highly inclined  ($85$ deg.) and an equatorial starting condition.
In the case of NRLMSISE-00, where the model lacks differentiability, we resort to the Dormand-Prince numerical scheme for numerical integration. However, for the fully differentiable model $\mathcal N_{NRL00}$, we opt for a Taylor integration method, implemented in the Heyoka open-source software \cite{biscani2022reliable}. The observed speedup (larger than one hundred, see Tab.~\ref{tab:speedup}) can be attributed to the combined advantages of the rapid inference time of thermoNETs and the utilization of the Taylor scheme \cite{biscani2022reliable}. To distinguish between these two effects, we report in the table one additional experiment measuring the propagation time using the Dormand-Prince scheme while evaluating the thermosphere density with the thermoNETs.

\begin{table*}[tb]
\begin{center}
\begin{tabular}{ c|c|c|c } 
 Numerical Technique & Thermosphere model & CPU time & speedup \\ 
  \hline
 Dormand-Prince & NRLMSISE-00 & 7.18s & x1\\ 
 Dormand-Prince & $\mathcal N_{NRL00}$ & 1.3s & x5.5 \\ 
 Taylor \cite{biscani2022reliable} & $\mathcal N_{NRL00}$ & 0.55ms & x130 \\ 
\end{tabular}
\end{center}
\caption{Indicative CPU timings (Intel(R) Xeon(R) CPU E5-2687W v3 @ 3.10GHz) for a single orbital propagation of a VLEO satellite ($t=50$ hours, $h=200$ km, $i=0$ deg.) undergoing a perfectly Keplerian motion perturbed by a drag term.\label{tab:speedup}}
\end{table*}

In Fig.~\ref{fig:orbits} we report the altitude trends during the propagation as computed for all cases considered. The vanishingly small error introduced when using the thermoNET is a further validation of the thermospheric models produced.

\subsection{NeuralODEs training}
\begin{figure}[tb]
  \centering
    \includegraphics[width=\columnwidth]{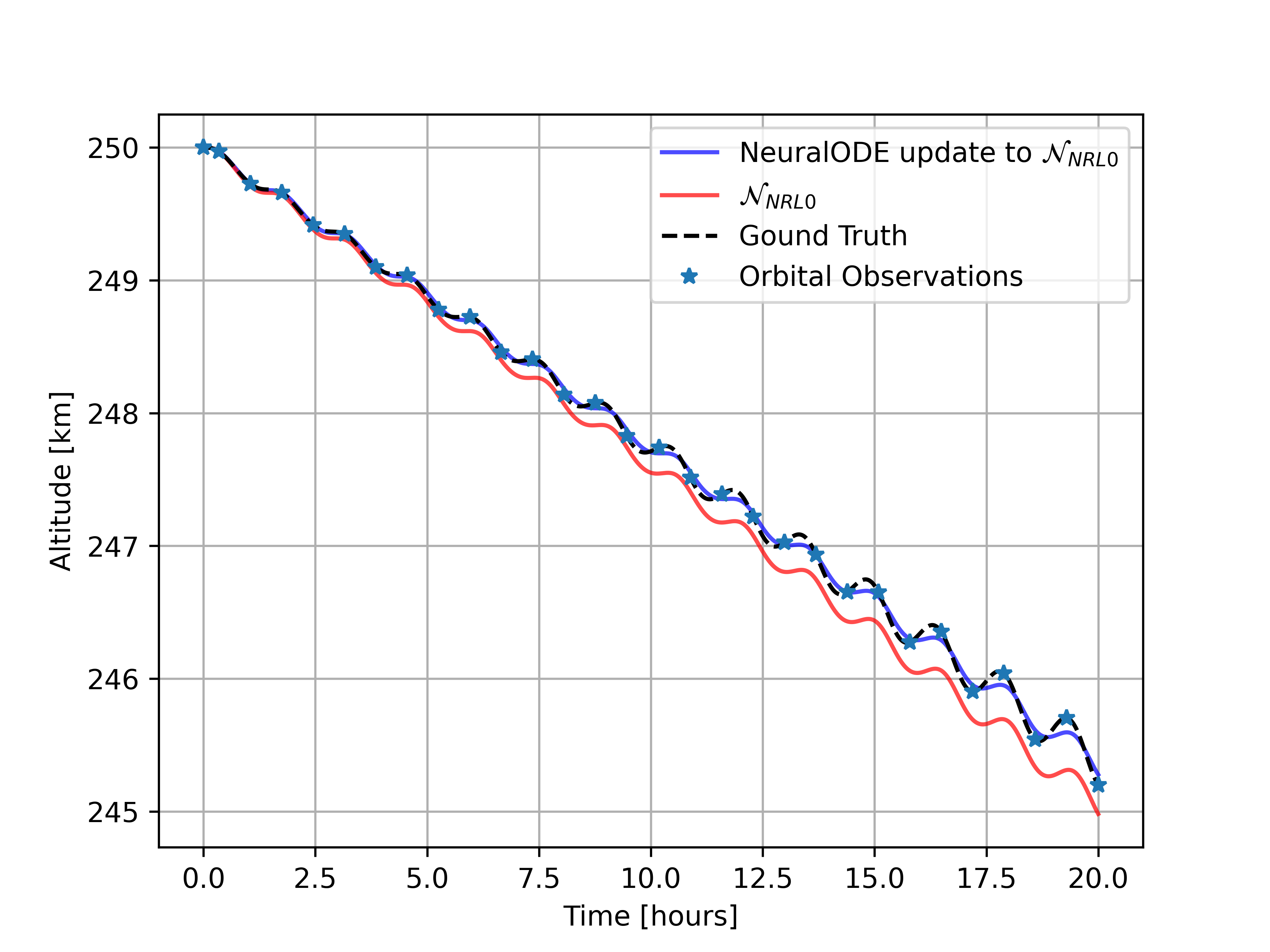}
  \caption{The updated thermospheric model learns to match orbital observations using the NeuralODE approach and captures the orbital decay rate accurately.
  \label{fig:neuralode}}
\end{figure}

We here use the term NeuralODE to indicate a set of Ordinary Differential Equations (ODEs) where the right hand side contains an artifical neural network. The term was coined in 2018 \cite{chen2018neural} and has since been widely adopted in various contexts. While there is not any formal difference between the theory of parametric ODEs and NeuralODEs, the presence of a neural network allows to use the universal approximation theorem \cite{tikk2003survey} to expect a great practical flexibility of the resulting system. In our case we consider the dynamics in Eq.~\eqref{eq:neuralODE} where the expression of the density $\rho = \rho_{\pmb \theta}$ is defined through a neural model parameterized by $\theta$ (e.g. weights and biases). Thanks to the differentiability of the neural model we derive, in an automated fashion, the variational equations and thus obtain the augmented system:
\begin{equation}
\label{eq:variational}
\left\{
\begin{array}{l}
\dot{\pmb r} = \pmb v \\
\dot{\pmb v} = \mathbf f(\pmb r, \pmb v, t) = -\frac{\mu}{r^3} \pmb r - \frac 12  \frac{\rho_{\pmb \theta}}{C_b} v_{r}\pmb v_r \\
\dot{\pmb \varphi} = \left[\frac{\partial \pmb f}{\partial \pmb r}, \frac{\partial \pmb f}{\partial \pmb v}\right]^T\pmb \varphi +  \frac{\partial \pmb f}{\partial \pmb \theta}
\end{array}
\right.
\end{equation}
where $\pmb \varphi = \left[\frac{\partial \pmb r}{\partial \pmb \theta}, \frac{\partial \pmb v}{\partial \pmb \theta}\right]^T$ is the gradient of the trajectory with respect to the ODE parameters and can thus be used to adjust them, via some form of gradient descent, and minimize a loss representing the deviation of the trajectories from observational data. 
To show the use of this technique (enabled practically by the small dimension of our thermoNETs and their differentiability), we simulate noiseless observations (on the full satellite state) of a ground truth VLEO orbit using a to-be-considered-unknown thermospheric model\footnote{For simplicity we use the JB08 empirical model when simulating a ground truth.}. We then update our $\mathcal N_{NRL0}$ thermoNET via a few steps of gradient descent aiming to minimise the loss between the synthetic observations and the results of the orbital propagation. The result, shown in Fig.~\ref{fig:neuralode}, shows how the updated thermospheric neural model captures accurately the ground truth orbital decay rate.

\section{Conclusion}
ThermoNETs are introduced as an innovative implicit neural representation of the density in the thermosphere, characterized by their reduced size, increased accuracy, and guaranteed asympthotic behaviour. We find they have the capability to approximate empirical models such as NRLMSISE-00 and JB-08 within a few percentage points of relative error across a wide altitude range and to offer a computationally efficient and easily manageable differentiable model. Through their combined use with techniques such as Taylor integration and NeuralODEs, ThermoNETs allow for substantial improvements in both propagation times and prediction accuracy for decay rates, particularly when trained with observational data.

\begin{figure}[tb]
  \centering
    \includegraphics[width=\columnwidth]{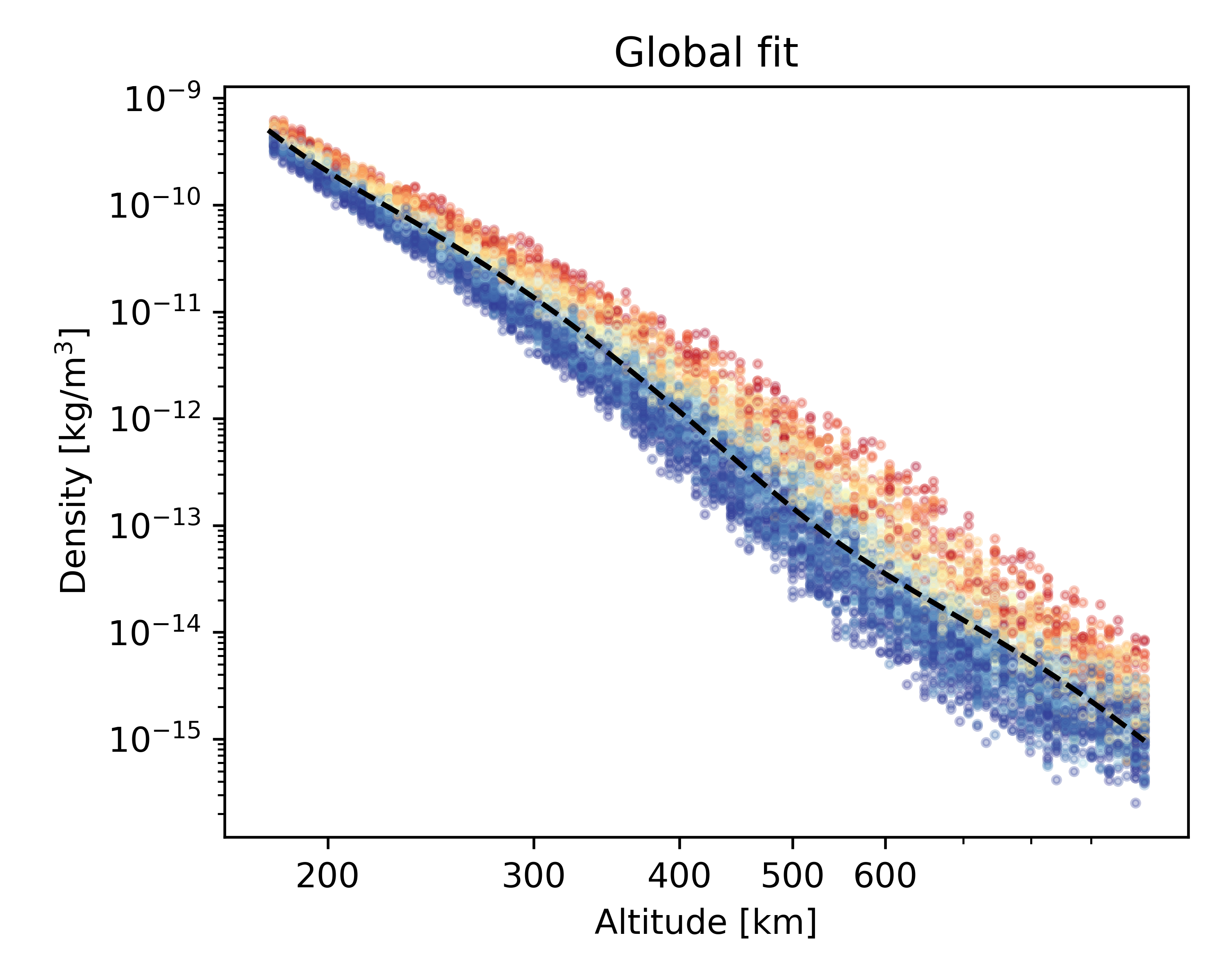}
  \caption{Global fit (dashed line) and the fitted NRLMSISE-00 data, differing in longitude, latitude and solar weather parameters. Color is indicating the $f_{10.7}$ coefficient with red being higher (higher solar activity). 
  \label{fig:rho_global_fit}}
\end{figure}

\section{Code availability}
All of the figures and quantitative results in this paper can be reproduced running the python based code distributed freely under a permissive GPLv3 license here: \url{https://github.com/esa/thermonets}. Refer to this resource to have all the numerical details on our simulations.

\bibliographystyle{ISSFD_v01}
\bibliography{references}

\clearpage
\section*{Appendix}

\appendix
\section{The Geodetic Coordinates}
\label{app:geodetic}
In order to compute the required thermoNETs inputs from satellite positions, we must invert the relations:
$$
\begin{array}{l}
x = (N+h) \cos\varphi\cos\lambda \\
y = (N+h) \cos\varphi\sin\lambda \\
z = \left((1-e^2)N+h\right) \sin\varphi
\end{array}
$$
where $e^2 = 1-\frac{b^2}{a^2}$, $N = \frac{a}{\sqrt{1 -e^2\sin^2\varphi}}$. $a$ and $b$ are the semi-major and semi-minor axis of the oblate ellipsoid the Geodetic coordinates refer to. The inverse relation, instead, can be written as:
$$
\begin{array}{l}
h = \frac {p}{\cos\varphi} - N \\
\varphi = \arctan \left(\frac z{p\left(1-e^2 \frac N{N+h}\right)} \right) \\
\lambda = \arctan2{(y,x)}
\end{array}
$$
where $p = \sqrt{x^2+y^2}$. 
These last relations require an iterative procedure in order to be solved as $\varphi$ appears on the right hand side explicitly via the definition of $N$. 
The discussion on how to compute geodetic coordinates from Cartesian coordinates has been profusely discussed in the literature over the past 40 years and can considered as a solved problem for our purposes (see \cite{featherstone2008closed} for a review of various approaches). 
We employ Algorithm \ref{alg:2geodetic} to perform the transformation. Note that we use a fixed number of iterations, namely four, which allow us to use the iterations to establish analytical and differentiable relations between the two sets of coordinates: $h(x,y,z), \varphi(x,y,z)$ and $\lambda(x,y,z)$. 
In the case of the Earth ($e=0.08181919$ from the WGS-84 ellipsoid \cite{slater1998wgs}), we find errors smaller than 70m when using 2 iterations, 40 cm when using 3 iterations, and 4mm when using 4 iterations.

\begin{algorithm}[tb]
\caption{Transforming Cartesian to Geodetic coordinates}\label{alg:2geodetic}

\begin{algorithmic}
\Require $n \geq 0$
\State $\lambda \gets \arctan2 (y,x)$
\State $p \gets \sqrt{x^2+y^2}$
\State $e2 \gets 1-\frac{b^2}{a^2}$
\State $\varphi \gets \arctan{\frac z {p(1-e^2)}}$
\For{$i = 1..4$} 
\State $N \gets \frac{a}{\sqrt{1 -e^2\sin^2\varphi}}$
\State $h \gets \frac p{\cos\varphi} - N$
\State $\varphi \gets \arctan{\frac z {p(1-e^2\frac{N}{N+h})}}$
\EndFor
\end{algorithmic}

\begin{tikzpicture}
\draw (0,0) ellipse (2.5cm and 1.5cm);
\draw[->] (-3,0) -- (3,0) node[right] {$\hat{\mathbf e}_1$};
\draw[->] (0,-2) -- (0,2) node[above] {$\hat{\mathbf e}_3$};
\coordinate (N) at (1, 1.375);
\coordinate (P) at (1.4, 2.7);
\coordinate (C) at (0.575, 0);
\coordinate (Cx) at (1, 0);
\filldraw (N) circle (1pt) node[above left] {$P'$};
\draw[dotted, blue] (N) -- (P) node[midway, right] {$h$};
\filldraw (P) circle (1.5pt) node[above, right] {$P$};
\filldraw (C) circle (1pt) node[below] {$C$};
\pic [draw, ->, blue, "$\varphi$", angle eccentricity=1.5] {angle = Cx--C--N};
\draw (C) -- (N) node[midway, right]{};
\draw[blue] [-latex, thick, rotate=90] (0.8,0.1) arc [start angle=10, end angle=350, x radius=0.15cm, y radius=0.4cm] node[above left] {$\lambda$};
\coordinate (Xm) at (0, -2.3);
\coordinate (XM) at (2.5, -2.3);
\draw[|<->|] (Xm)--(XM) node[below]{$a$};

\coordinate (Ym) at (-3.3, 0);
\coordinate (YM) at (-3.3, 1.5);
\draw[|<->|] (Ym)--(YM) node[left]{$b$};
\end{tikzpicture}


\end{algorithm}

\section{Training details}
\label{sec:training}
For both networks, $\mathcal N_{NRL00}$ and $\mathcal N_{JB08}$, we use as inputs the two geodetic coordinates $\lambda, \varphi$ (the altitude is taken care of by Eq.~\eqref{eq:rho_functional_form}), the day of year (DOY) the seconds in day (SID) (a proxy for the Local Solar Time) and the relevant solar weather parameters $\mathcal S_w$. Similarly to what is done in \cite{licata2021improved} we avoid angular discontinuities by augmenting the attributes as follows:
$$
t_1 = \sin\left(\lambda \right), t_2 = \cos\left(\lambda \right)
$$
$$
t_3 = \sin\left(2\pi \frac{DOY}{365.25} \right), t_4 = \cos\left(2\pi \frac{DOY}{365.25} \right)
$$
$$t_5 = \sin\left(2\pi \frac{SID}{86400} \right), t_6 = \cos\left(2\pi \frac{SID}{86400} \right)
$$
In the case of $\mathcal N_{NRL00}$ the space weather indices used as attributes are $\mathcal S_w = [f_{10.7}, f_{a10.7}, A_p]$, representing the solar radio flux at 10.7 cm ($f_{10.7}$), its 81 days average ($f_{a10.7}$) and an average level for the geomagnetic activity ($A_p$). For $\mathcal N_{JB08}$ the space weather indices considered are instead those used in the Jacchia-Bowman empirical model and hence include more solar flux descriptors $[f_{10.7}, s_{10.7}, m_{10.7}, y_{10.7}]$, their 81 day averages as well as the geomagnetic indicators $A_p$ and $Dst$. The network architecture uses $\tanh$ non-linearities and include, after the input layer, two hidden layers of 32 neurons each. The thermoNETs outputs 12 quantities $\delta\alpha_i, \delta\beta_i,\delta\gamma_i, i=1..4$ which reconstruct a density value using Eq.\eqref{eq:rho_functional_form}, where:
\begin{equation}
\begin{array}{l}
\alpha_i = \overline \alpha_i (1 + \delta\alpha_i)\\
\beta_i = \overline \beta_i (1 + \delta\beta_i)\\
\gamma_i = \overline \gamma_i (1 + \delta\gamma_i)\\
\end{array}
\end{equation}
The reference values $ \overline \alpha_i, \overline \beta_i, \overline \gamma_i$ are obtained fitting Eq.~\eqref{eq:rho_functional_form} the whole dataset. In Fig.~\ref{fig:rho_global_fit} we report the trend of such a global fit (only dependent on the altitude) used to predict the density values in the $\mathcal D_{NRL00}$ dataset. The networks inputs are min-max scaled and the parameters are trained using the Adam optimizer with a learning rate scheduler.

\end{document}